\documentclass{ws-p8-50x6-00}
\usepackage[dvips]{color}       
\usepackage{epsfig}

\def\pmb#1{\setbox0=\hbox{#1}
\kern-.025em\copy0\kern-\wd0
\kern.05em\copy0\kern-\wd0
\kern-.025em\raise.0433em\box0}
\def\mbi#1{{\pmb{\mbox{\scriptsize ${#1}$}}}}

\def\bm#1{{\pmb{\mbox{${#1}$}}}}
\def\strut{\vrule width0pt height 15pt depth 7pt}

\begin{document}

\def\beq{\begin{equation}}
\def\eeq{\end{equation}}
\def\beqa{\begin{eqnarray}}
\def\eeqa{\end{eqnarray}}
\def\d{{\rm d}}
\def\ttimes{{\scriptstyle \times}}
\def\half{{\textstyle {1\over2}}}

\title{RADIAL EXCITED STATES OF THE NUCLEON IN QUARK MODELS WITH DYNAMICAL CONFINEMENT}

\author{M. Fiolhais and P. Alberto}

\address{Departamento de F\'\i sica and Centro de F\'\i sica Computacional, 
Universidade de Coimbra, P-3004-516 Coimbra, Portugal\\ 
E-mail: tmanuel@teor.fis.uc.pt}

\author{B. Golli and S. \v{S}irca}

\address{University of Ljubljana and J. Stefan Institute, Ljubljana, Slovenia}

\maketitle

\abstracts{
The Roper state is described in the framework of the chiral
chromodielectric model as a single quark excitation in self-consistent
meson potentials.  The interplay between quark and meson excitations
is discussed and we emphasise the role of non-quark degrees of freedom
in the electroproduction of nucleon excitations.}

\section{Introduction}

The $P_{11}(1440)$ (Roper) resonance~\cite{roper} is the lowest
positive-parity $\mathrm{N}^\star$ state, most clearly visible
in partial-wave decompositions of
$\pi\mathrm{N}\to\pi\mathrm{N}$ and $\pi\mathrm{N}\to\pi\pi\mathrm{N}$
scattering, as well as in pion photoproduction.
Although this four-star resonance is within the energy reach
of many modern research facilities, the experimental analyses
so far have not ventured beyond the determination of its mass,
width, and electromagnetic couplings~\cite{pdg}.  The only
experimental information on the helicity amplitudes for Roper
excitation with virtual photons comes from a very scarce data
set obtained in older electroproduction experiments at DESY
and NINA~\cite{gerhardt,boden}.

The electroexcitation of the Roper is also receiving considerable
theoretical interest. In the spherically-symmetric $\mathrm{SU}(6)$
quark model, the Roper can be understood as a radial excitation
(``breathing mode'') of the proton, with one quark occupying
the $2\mathrm{s}$ state, yielding a $(1\mathrm{s})^2(2\mathrm{s})^1$
configuration.  This physical picture implies a sizable Coulomb
monopole contribution.  Newer developments~\cite{li1,li2} have
indicated a possible description of the Roper as a gluonic partner
of the proton, represented as a $(\mathrm{q}^3\mathrm{g})$
hybrid baryon.  In this approach, the monopole strength 
is expected to be highly suppressed, in contrast to the
concept of ``breathing''.  These two opposing concepts result
in very different predictions for the $Q^2$-dependence
of the transverse ($A_{1/2}$) and scalar ($S_{1/2}$)
electroproduction helicity amplitudes, and they both fail
to reproduce the electromagnetic couplings in the real-photon limit.
The Roper has been investigated as a hybrid baryon by using
QCD sum rules~\cite{kiss}, and in terms of vibrating flux-tubes
between quarks~\cite{capstick4,strobel2}.
Extensive studies in constituent, non-relativistic,
relativised, and relativistic quark models also 
exist~\cite{stancu}$^-$\cite{cardarelli}.
They generally result in good descriptions of the mass and width,
but fail to reproduce the photo-couplings, with an exception
of a non-relativistic quark model extended with vector-meson
exchange \cite{cano}.  An important addition to quark models are
the mesonic degrees of freedom. A fair understanding of the
photo-couplings was obtained by introducing meson-exchange currents
between quarks \cite{meyer}, and in a relativistic quark model
with a three-quark core and an admixture of a pion-baryon
configuration~\cite{dong}.

In this work we use the framework of the chiral chromodielectric model 
to describe the nucleon and the Roper resonance as 
clusters of three valence quarks, coupled to clouds of $\sigma$
and $\pi$ mesons and to a chromodielectric field which dynamically
confines the quarks\cite{alberto,broRPA}.  The quarks are in
the $(1\mathrm{s})^3$
configuration for the nucleon and $(1\mathrm{s})^2(2\mathrm{s})^1$
for the Roper.  The physical Roper emerges after the radial fields
are determined in an iterative self-consistent variation using
angular momentum and isospin projections.

In section 2 we introduce the model and in section 3 we explain how model 
states representing the nucleon and the Roper are obtained. The role of
non-quark degrees of freedom in the Roper is discussed in section 4.
In section 5 we present the electromagnetic form factors
for the nucleon--Roper transition. 

\section{The model}\label{model}

The CDM contains quark and chiral meson degrees of freedom, 
in addition to a scalar-isoscalar chiral singlet chromodielectric field. 
The coupling of this field to fermions leads to quark
confinement, and is an important feature of the model.
The Lagrangian of the model reads~\cite{neuber}
\begin{eqnarray}
{\mathcal L} &=& {\rm i} \, \overline q \gamma^\mu\partial_\mu q + 
\frac{g}{\chi}
\overline q \left(\sigma_0 + {\rm i} \vec\tau\cdot\vec\pi \gamma_5\right) q +
\frac{1}{2}\left(\partial_\mu \sigma_0\right)^2+
\frac{1}{2}\left(\partial_\mu\vec\pi\right)^2+
\frac{1}{2}(\partial_\mu \chi)^2  \nonumber \\
& & -W(\sigma_0,\vec\pi) - U(\chi)\, .
\label{eq:1}
\end{eqnarray}
Here $q(x)$ stands for the quark
field operator, $\vec{\pi}(x)$ and $\sigma_0(x)$ are the
chiral meson fields, pion and sigma,  respectively
(the arrow denotes isovector), and $\chi(x)$ is the
chromodielectric field. There are two potential terms in
Eq.~(\ref{eq:1}): ${W}(\sigma,\vec\pi)$, which
is the Mexican hat potential for the chiral fields~\cite{neuber,BirseP},
and $U(\chi)$, the $\chi$ field potential. The Mexican hat potential is 
\beq
W={\lambda\over 4} \left( \sigma_0^2 + \vec \pi ^2 -\nu^2  \right) ^2 +
c \sigma_0 + W_0\, ,
\label{xpmh}
\eeq
where the parameters $\lambda$, $\nu$, c and $W_0$ are related to the 
chiral meson masses and to the pion decay constant:
\beq
\lambda={m_\sigma^2 -m_\pi^2 \over 2 f_\pi^2}\, , \ \ \ \nu= f_\pi^2 - 
{m_\pi^2 \over \lambda}\, ,  \ \ \ c= - f_\pi m_\pi^2 \, .
\eeq
The constant $W_0$ in Eq. (\ref{xpmh}) ensures that min$(W)=0$. For the
 $\chi$ field potential we take  
\begin{equation}
U(\chi)= {1 \over 2} M^2 \chi^2 
\label{uchi}
\end{equation}
where $M$ is the mass of the field. 

The vector current is conserved for the Lagrangian (\ref{eq:1}), which is  
almost chiral SU(2)$\times$SU(2) symmetric. It is the (small) linear sigma term
in (\ref{xpmh}) that explicitly breaks the chiral symmetry of (\ref{eq:1}), 
thus enforcing PCAC. 
The Mexican hat potential induces spontaneous chiral symmetry breaking, 
the vacuum expectation values of the chiral meson
being $\langle 0 | \vec\pi | 0 \rangle =0$ and 
$\langle 0 | \sigma_0 | 0 \rangle = -f_\pi$, and a (dynamical)
mass is acquired by the fermions.   The parameters in the Mexican
hat potential are the pion mass ($m_\pi=0.14$~GeV), the pion decay
constant ($f_\pi=0.093$~GeV), and the sigma mass ($m_\sigma=0.85$~GeV).
In addition, there is one more parameter in model (\ref{eq:1}),
the coupling constant $g$. For quadratic potentials such as (\ref{uchi}), 
it turns out that the evaluated properties of the model states
depend on $G = \sqrt{gM}$~\cite{drago2}, but not on $g$ or $M$ separately.

The CDM describes a system of interacting quark and meson fields. 
In the baryon sector of the
model, one finds soliton solutions with three confined quarks due to  
the peculiar way the $\chi$ field couples to the fermions, which leads to 
an $r-$dependent quark mass, $m_q(r)={g f _\pi /  \chi(r)}$. The soliton 
requires $\chi(r)\rightarrow 0$ as $r\rightarrow \infty$, and hence
$m_q(r)\rightarrow\infty$.  This mechanism of increasing mass
prevents the quarks to move too far away from the origin. 
The $\chi$ field can be regarded as a 0$^{++}$ glueball~\cite{BirseP}. 
It is an effective field mimicking the confining mechanism of QCD.

\section{Baryons in the CDM model}\label{baryons}
 
Coherent states are assumed for all mesons and the starting point
to describe a  baryon is the hedgehog coherent state around a cluster
of three valence quarks, which we write in the form
\begin{eqnarray}
|Hh\rangle &=& N \, {\rm exp} \left\{ \sum_{tm} (-1)^{1-m} \delta_{t,-m}
\int_0^\infty \d k \sqrt{2 \pi \omega_\pi(k) \over 3} \xi(k)
a^\dagger_{tm} (k) \right\}\times     
\nonumber \\ &&    {\rm exp}   
\left\{ \int_0^\infty \d k \sqrt{2 \pi \omega_\sigma(k) } \eta(k)
\tilde{a}^\dagger (k)\right\}\times     \nonumber \\ &&
{\rm exp}\left\{ \int_0^\infty \d k \sqrt{2 \pi \omega_\chi(k) }
\zeta(k) {b}^\dagger (k)\right\}
\times \prod_{i=1,3} c^\dagger_h (i) |0\rangle \, ,
\label{eq:3}
\end{eqnarray}
\vskip0.2cm 
\noindent where $N$ is a normalization constant. The pion, the sigma and the chi
amplitudes, $\xi(k)$, $\eta(k)$ and $\zeta(k)$, are related to the pion,
sigma and chi radial profiles through
\beq
\phi(r)= \sqrt{2 \over \pi} \int_0^\infty \!\!\! k^2 \d k j_1(kr)  \xi(k) \, ,
\ \ \ \
\left\{
\begin{array}{c}
\sigma(r) \\ \chi(r) 
\end{array} 
\right\}
= \sqrt{2 \over \pi} \int_0^\infty \!\!\! k^2 \d k j_0(kr)  
\left\{
\begin{array}{c}
\eta (k)\\ \zeta (k) 
\end{array} 
\right\} \, . \label{radiif}
\eeq 
\vskip0.2cm 
The expectation values of the meson field operators in the hedgehog state are
\beq
\langle \pi_t({\bm r})\rangle
=\sqrt{4\pi \over 3 } Y^*_{1t}(\hat{\bm r}) \, \phi(r)  \, , 
\langle\sigma({\bm r})\rangle= \sigma(r) \, ,
\langle\chi({\bm r})\rangle= \chi(r)  
\label{xmesons}
\eeq
\vskip0.2cm 
\noindent ($t=0,\pm 1$ is the third component of isospin). In Eq. (\ref{eq:3}),
$a^\dagger_{tm}$ is the creation operator for a p-wave pion with
third components of isospin and angular momentum $t$ and $m$, respectively,
orbital wave function $\xi(k)$, and frequency $\omega_\pi=\sqrt{k^2+m_\pi^2}$.
Similarly, $\tilde{a}^\dagger (k)$ and ${b}^\dagger (k)$ create sigma
and chi quanta in s-wave, with orbital wave functions $\eta(k)$
and $\zeta(k)$, respectively, and frequencies $\omega_\sigma$
and $\omega_\chi$.  Finally, the operator  $c_h^\dagger(i)$
creates a s-wave valence quark in a spin-isospin hedgehog state:
\vskip0.2cm 
\beq
\langle{\bm r} | c^\dagger_h(i) |0\rangle = q_i({\bm r})=
{1 \over \sqrt{4 \pi}} \left( 
\begin{array}{c}
   u_i(r)  \\ {\rm i} v_i(r) \, {\bm \sigma} \cdot \hat{\bm r}
\end{array}
\right)
|h\rangle \, , \ \  |h\rangle={1 \over \sqrt{2}}
\left( |u \downarrow \rangle - | 
d \uparrow \rangle \right) \, . \label{eq:::10}
\eeq
\vskip0.2cm 
\noindent The index $i$ stands for all quantum numbers and distinguishes
between different radial states.  
Quarks are always assumed to be in an s state ($j=1/2$). Hence,
only p-wave pions, and s-wave sigmas and chis may take part in the
interactions.  

Because (\ref{eq:3}) is not an eigenstate of angular momentum
or isospin, the representative model states 
of the nucleon and Roper are obtained through a standard
Peierls-Yoccoz projection on angular momentum (and isospin) \cite{GR85}, 
\vskip0.2cm 
\beq
|J,M;T=J, M_T\rangle
= {1 \over \sqrt{F_J}} (-1)^{J+M} P^J_{M, -M_T} |Hh\rangle
\label{eq:6}
\eeq  
where 
\beq
P^J_{M,K} = {2 J + 1 \over 8 \pi^2} \int \d \Omega \, \,
{\mathcal D}^{J^*}_ {M,K} (\Omega)\, R(\Omega)\, . \label{projector}
\eeq
Here $R(\Omega)$ is the rotation operator and ${\mathcal D}$ are
the Wigner functions. In (\ref{eq:6}) $F_J$ is the normalization
factor given by 
\beq
F_J= \langle Hh | P^J_{KK} | Hh \rangle \, , \label{fcorpo}
\eeq
which does not depend on $K$ (so we choose $K=J$). 
For both the nucleon and the Roper, $J=T={1\over 2}$.  Due to the 
symmetry properties of the hedgehog, only a single projection,
either in angular momentum or in isospin, is required since one
operation in either space automatically selects the subspace
$T=J$ in the other space.  In Eq. (\ref{eq:6}) we project
onto  angular momentum. 

The radial profiles $\phi(r)$, $\sigma(r)$, $\chi(r)$, $u_i(r)$, and
$v_i(r)$ in Eqs. (\ref{xmesons}) and (\ref{eq:::10}) are determined
using (\ref{eq:6}) as a trial state,  applying a variational principle
to the angular momentum--isospin ``projected energy", 
$E_{J}=\langle JM,TM_T | H_{\rm CDM} | JM,TM_T\rangle$, which is given by
\vskip0.2cm 
\beqa
&&\!\!E_{J}\!\!=\sum_{i=1}^3 \!\epsilon_i +\!\!
4 \pi \int_0^\infty \!\!\!r^2 \d r \left\{
 \left[ \left({\partial \sigma \over \partial r} \right)^2 +
m_\sigma^2 \sigma ^2\right] \!\! + \!\!
\left[ \left({\partial \phi \over 	\partial r} \right)^2  +
{2 \phi^2\over r^2}+ 
m_\pi^2 \phi ^2\right] {\cal C}^J_0 \right. \nonumber \\
&&\!\!\! \left. +\left[ \left({\partial \chi \over \partial r} \right)^2
+ M^2 \chi ^2\right]
+{\lambda\over 4} \left[\phi^4 {\cal C}^J_4 +
2 \phi^2 (\sigma^2 -2 f_\pi \sigma) {\cal C}^J_2
+\sigma^4-4 f_\pi \sigma^3\right] \right\}\! . \ \label{energy}
\eeqa
\vskip0.2cm 
The ``projection coefficients" ${\cal C}^J$ are expressed in terms of the 
normalization functions (\ref{fcorpo}) which, in turn, depend on the 
intrinsic number of pions 
\beq
N_\pi = 4 \int_0^\infty \d k k^2 \omega_\pi(k)  \left[\int_0^\infty  \d r r^2 
\phi(r) j_1(kr)\right]^2\, . \label{number0}
\eeq
In (\ref{energy}), $\epsilon _i$ are quark energy eigenvalues. They are all 
equal for the nucleon. For the Roper two of them are equal (corresponding
to the lowest state, 1s), while the third one corresponds to the 2s state.

The nucleon and the Roper resonance are therefore described
as clusters of three quarks in radial-orbital configurations
(1s)$^3$ and (1s)$^2$ (2s)$^1$ respectively, each surrounded
by pion and $\sigma$-meson clouds and by a chromodielectric field,
projected onto subspace with good angular momentum and isospin:
\begin{equation}
|{\rm N}_{{1\over 2},M_T}\rangle 
   = \mathcal{N}\,
     P^{1\over 2}_{{1\over 2},-M_T} |Hh\rangle \, , 
\quad 
    |{\rm R}'_{{1\over 2},M_T}\rangle 
     = \mathcal{N}'\,
     P^{1\over 2}_{{1\over 2},-M_T} |Hh^*\rangle 
\label{usa}
\end{equation}
(${\cal N}$ and ${\cal N}'$ are normalization factors).
The Roper and the nucleon have different boson fields profiles since 
they can adapt to the corresponding quark configuration.
The proper orthogonalization of  states (\ref{usa}) is ensured 
by writing: 
\begin{equation}
 |\tilde{\rm R}\rangle = c^R_R|{\rm R}'\rangle + c^R_N|{\rm N}\rangle\,, 
\quad
 |\tilde{\rm N}\rangle = c^N_R|{\rm R}'\rangle + c^N_N|{\rm N}\rangle 
\label{gcmss}
\end{equation}
where the coefficients result from the diagonalization of the
Hamiltonian in the subspace spanned by $|{\rm R}'\rangle$
and $|{\rm N}\rangle$. However, we verified that the 
simpler orthogonalization procedure 
\begin{equation}
   |{\rm {R}}\rangle 
    = {1\over\sqrt{1-c^2}}(|{\rm {R}}'\rangle - c|{\rm {N}}\rangle)\;,
\qquad
  c = \langle{\rm {N}}|{\rm {R}}'\rangle 
\label{orthoR}
\end{equation}
is good enough. 

A central point in our treatment of the Roper 
is the freedom of the chromodielectric profile, 
as well as of the chiral meson profiles, 
to adapt to the (1s)$^2$(2s)$^1$ configuration. 
Therefore, quarks in the Roper experience mean fields 
which are different from the mean boson fields felt by the quarks 
in the nucleon. In other words, the `potential' breathes together with the quarks
as illustrated in Figures~1 and 2. 

\begin{figure}[hb]
\centerline{\epsfig{clip=on,file=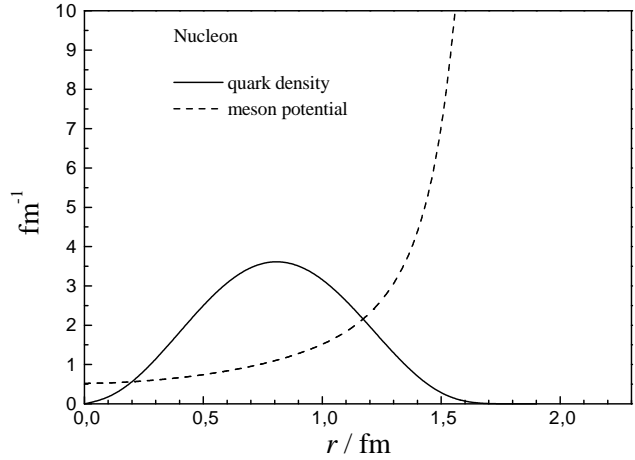,width=9.5cm}}
\caption{The baryon density
(solid line) and the effective potentials (dashed line) generated 
by the self-consistently determined $\pi$, $\sigma$ and $\chi$ fields
in the nucleon.}
\end{figure}

\begin{figure}[ht]
\centerline{\epsfig{clip=on,file=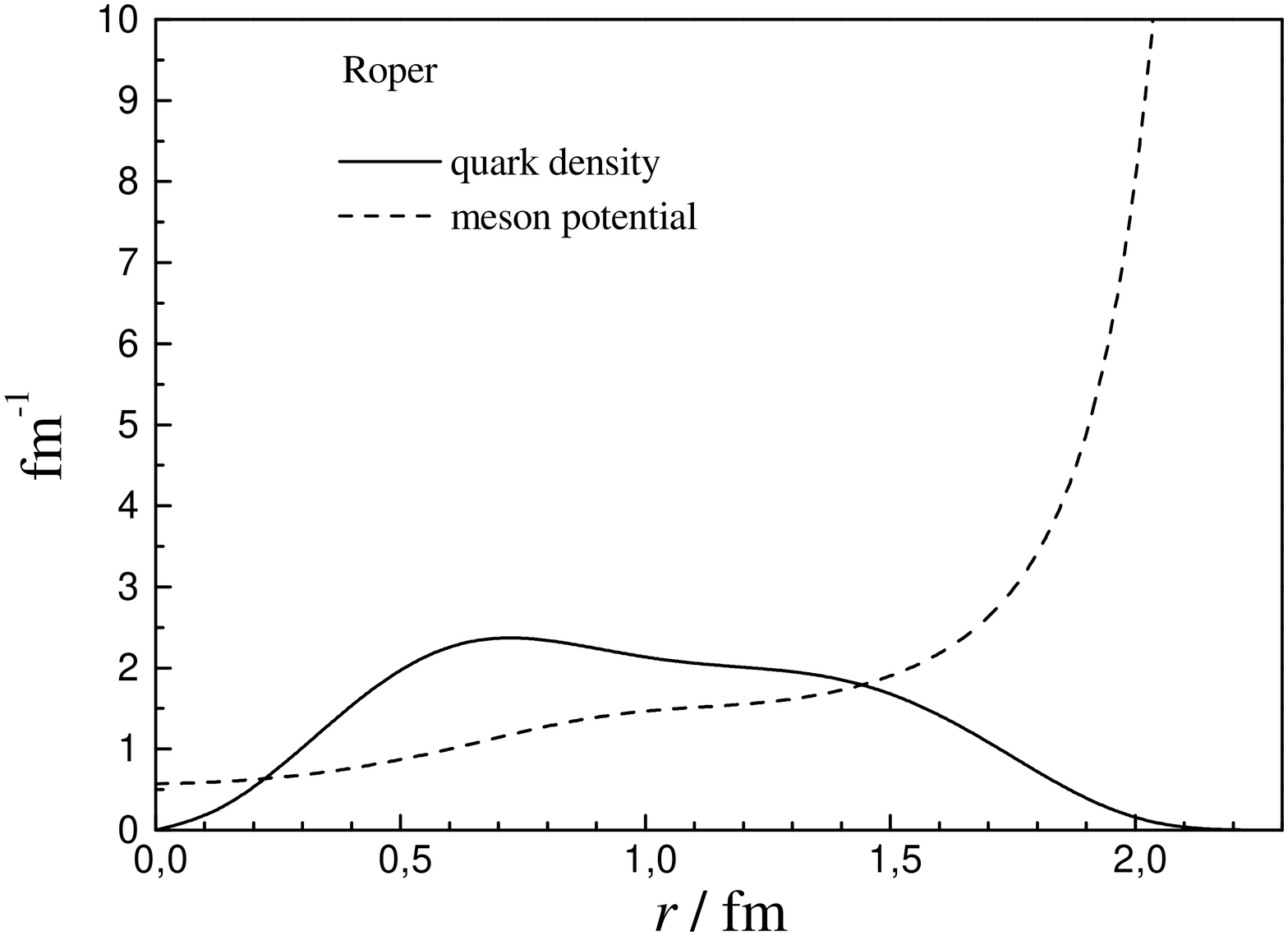,width=9.5cm}  }
\caption{The baryon density
(solid line) and the effective potentials (dashed line) generated 
by the self-consistently determined $\pi$, $\sigma$ and $\chi$ fields
in the Roper.}
\end{figure}

\section{Non-quark degrees of freedom}

We have also investigated another possible type of excitation in which 
the quarks remain in the ground state configuration (1s)$^3$ while 
the chromodielectric field and the $\sigma$-field oscillate. 
Such oscillations can be described by expanding the boson fields as 
small oscillations around their ground-state values.
For the $\sigma$-field we write:
$$
  \hat{\sigma}(\vec{r}) =
     \sum_n{1\over\sqrt{2\varepsilon_n}}\, \varphi_n(r)
     {1\over\sqrt{4\pi}}
     \left[{\tilde{a}}_n + {\tilde{a}}^\dagger_n\right] + \sigma(r)\, ,
$$
where
$\varphi_n$ and $\varepsilon_n$
satisfy the Klein-Gordon equation
\begin{equation}
 \left(-\nabla^2 + m^2 
     + {\d^2 {V}({\sigma}(r))\over\d{\sigma}(r)^2}\right)\varphi_n(r)
  = \varepsilon_n^2 \varphi_n(r)\;.
\label{KG}
\end{equation}
A simple ansatz for the anni\-hi\-la\-tion 
(creation) operator of the $n$-th mode is given by\footnote{This operator is not 
to be confused with the free field operator $\tilde a$ introduced in (\ref{eq:3}).}
\begin{equation}
 {\tilde{a}}_n = \int\d k\, \tilde{\varphi}_n(k)
    \left({\tilde{a}}(k) 
    - \sqrt{2\pi\omega_\sigma(k)}\,\eta(k)\right)\;,
\qquad
{\tilde{a}}_n|N\rangle = 0\;,
\label{an}
\end{equation}
where $\eta(k)$ and $\tilde{\varphi}_n(k)$ are the Fourier 
transforms of $\sigma(r)$ and $\varphi_n(r)$, respectively, 
and $\omega_\sigma(k)=\sqrt{k^2+m_\sigma^2}$. 
The effective potential in (\ref{KG}) is given by
\begin{equation}
 V_{\sigma\sigma}(r) 
  = \lambda \left[C_2\phi(r)^2 
        + 3\sigma(r)(\sigma(r)+2\sigma_v)\right] \, ,
\label{Vsigma}
\end{equation} 
where $\sigma$ is the fluctuating part of the full field and
$C_2$ is a projection coefficient slightly smaller than unity.
Similar expressions hold for the $\chi$ field with
the effective potential
\begin{equation}
 V_{\chi\chi}(r) 
  = -{3\over2\pi}\, {g\over\chi(r)^3}
   \left[(\sigma(r)+\sigma_v)(u(r)^2-v(r)^2) 
        + 2\phi(r)u(r)v(r)\right]\, .
\label{Vchi}
\end{equation} 
The effective potential turns out to be repulsive for the $\chi$-field 
and attractive for the $\sigma$-field; in the latter case there exists 
at least one bound state with the energy $\varepsilon_1$ of typically 
100~MeV below the $\sigma$-meson mass.

The ansatz for the Roper can now be simply extended as
\begin{equation} 
  |{\rm {R}}^*\rangle = c_1|{\rm {R}}\rangle 
       + c_2{\tilde{a}^\dagger}_\sigma|{\rm N}\rangle\;,
\label{Rext}
\end{equation} 
where $\tilde{a}^\dagger_\sigma$ is the creation operator 
for the lowest vibrational mode.
The coefficients $c_i$ and the energy are determined by solving 
the generalized eigenvalue problem in the $2\times2$ subspace.
The solution with the lowest energy corresponds to the Roper,
while the orthogonal combination to one of the higher excited 
states with nucleon quantum numbers, e.g., the ${\rm N}(1710)$, provided 
the $\sigma$-meson mass is sufficiently small.
The energy of the Roper is reduced (see Table~\ref{Tab:1}),
though the effect is small due to  weak coupling between the 
state (\ref{orthoR}) and the lowest vibrational state with 
the energy $\varepsilon_1$.
The reduction becomes more important if the mass of the
$\sigma$-meson is decreased.
The energy of the combination orthogonal to the ground state 
is close to $E_\mathrm{N}+\varepsilon_1$ with
$\sigma$-meson vibrational mode as the dominant component.

\begin{table}[h]
\begin{center}
\begin{tabular}{rrrrrrr}
\hline
\hline
\strut $m_\sigma$ & $\ \ \ \ E_\mathrm{N}$ \ & \ \ 2s--1s& 
\ \ \ \ $\Delta E_\mathrm{R}$& \ \ \ \ $\Delta E_{\mathrm{R*}}$& 
\ \ \ \ \ \  $c_2$& $\ \ \ \ \ \ \varepsilon_1$ \\ 
\hline
\strut 1200 & 1269 & 446 & 354  & 353 & 0.05 & 1090 \\
\strut  700 & 1249 & 477 & 367  & 364 & 0.12 &  590 \\
\hline
\hline 
\end{tabular}
\end{center}
\caption{For two $\sigma$-masses we show the nucleon energy
($E_\mathrm{N}$), the Roper-nucleon energy splittings calculated
from the single particle energy difference (2s--1s), the state
(\ref{orthoR})  ($\Delta E_\mathrm{R}$) and the state (\ref{Rext})
($\Delta E_{\mathrm{R*}}$). All energies are given in MeV.}
\label{Tab:1}
\end{table}

\section{Nucleon--Roper electromagnetic transitions}

The electromagnetic nucleon-Roper transition amplitudes
as well as the transition amplitudes to higher excitations with
nucleon quantum numbers represent an important test which may 
help distinguish between the models listed at the beginning. 
The transverse helicity amplitude is defined as
\begin{equation}
A_{1/2}=-{\zeta}\,\sqrt{2 \pi \alpha \over k_W}  \int \d^3 {\bm r} \, \
  \langle\tilde{\rm {R}}_{+{1\over 2}, M_T}|{\bm J}_{\rm em}({\bm r})\cdot 
   {\bm \epsilon}_{+1} \, {\rm e} ^{{\rm i}{\mbi k} 
     \cdot {\mbi r} } |\tilde{\rm {N}}_{-{1\over 2}, M_T }\rangle
\label{Ahalf}
\end{equation}
where $k_W$ is the photon momentum at the photon point,
and the scalar helicity amplitude as
\begin{equation}
S_{1/2}={\zeta}\,\sqrt{2 \pi \alpha \over k_W}  \int \d {\bm r}  \, 
    \langle\tilde{\rm {R}}_{+{1\over 2}, M_T}|{J}^0_{\rm em} ({\bm r}) 
      \,\, 
     {\rm e} ^{{\rm i}{\mbi k} 
     \cdot {\mbi r} } |\tilde{\rm {N}}_{+{1\over 2}, M_T}\rangle\, .
\label{Shalf}
\end{equation}
Here ${J}^\mu_{\rm em}$ is the EM current derived from the
Lagrangian density (\ref{eq:1}):
\begin{equation}
  {J}^\mu_{\rm em} = 
    \sum_{i=1}^3 {\overline{q}}_i \gamma^{\mu}(i) \left( {1 \over 6} 
  + {1 \over 2} \tau_0{(i)} \right) {q}_i +
   \left(\vec{{\pi}} \times \partial ^\mu \vec{{\pi}} \right)_0 \;.
\label{Jem}
\end{equation}
The amplitudes (\ref{Ahalf}) and (\ref{Shalf}) contain a phase factor
$\zeta$ determined by the sign of the decay amplitude into the nucleon 
and the pion.

The new term in (\ref{Rext}) {\em does not contribute}
to the nucleon-Roper transition amplitudes.
Namely, for an arbitrary EM transition operator $\hat{O}$ involving
only quarks and pions we can write
$ 
  \langle N|\tilde{a}_1 \hat{O}|N\rangle =
  \langle N|[\tilde{a}_1, \hat{O}]|N\rangle  = 0,  
$ 
because of (\ref{an}) and since the operators $\tilde{a}_n$ commute with 
$\hat{O}$. The amplitudes (\ref{Ahalf})--(\ref{Shalf}) obtained
in the framework 
of the present calculation can be found in Ref.~\cite{alberto}.  
We found a relatively  large discrepancy at the photon point which can be
attributed to a too weak pion field in the model.
Such small contribution was already noticed in the calculation 
of nucleon magnetic moments~\cite{drago2} and
of the electroproduction of the $\Delta$~\cite{delta1}.
The pion contribution to the charged states only accounts for 
a few percent of the total amplitude. 

Though the chromodielectric model 
gives only a qualitative picture of the lowest radially excited
states of the nucleon and their electroproduction amplitudes,
it yields some interesting features, in particular the possibility
of $\sigma$-meson vibrations.
Its main advantage over other approaches is that all properties,
including the electro-magnetic amplitudes and the resonance decay,
can be calculated from a single Lagrangian without additional
assumptions.  It also allows us to treat exactly  
the orthogonalization of states which is particularly important
in the description of nucleon radial excitations.

\vskip0.3cm 

This work was supported by FCT (POCTI/FEDER), Portugal, and by 
The Ministry of Science and Education of Slovenia.
MF acknowledges a travel  grant from Funda\c c\~ao Calouste
Gulbenkian (Lisbon), which made possible his participation
in Hadrons 2002. He also would like to thank the organizers
for the fine hospitality he enjoyed in Bento Gon\c calves.

\end{document}